\newcommand{\be}{\begin{equation}}\newcommand{\ee}{\end{equation}}
\newcommand{\bea}{\begin{eqnarray}}\newcommand{\eea}{\end{eqnarray}}
\newcommand{\nn}{\nonumber\\}\newcommand{\p}[1]{(\ref{#1})}
\newcommand{\supa}{superparticle }\newcommand{\sust}{superstring }
\newcommand{\susy}{supersymmetry }\newcommand{\sudi}{superdiffeomorphism }
\newcommand{\suli}{superworldline }\newcommand{\sush}{superworldsheet }
\documentstyle[12pt]{article}

\begin{document}
\renewcommand{\thefootnote}{\fnsymbol{footnote}}
\thispagestyle{empty}
{\hfill  Preprint JINR E2-92-337}\vspace{2.5cm} \\
\begin{center}
{\large\bf TWISTOR-LIKE TYPE II SUPERSTRING AND BOSONIC STRING} \vspace{1.5cm}
 \\
V. Chikalov \vspace{1cm} \\
{\it Institute of Applied Physics} \\
{\it 244007 Sumy, Ukraine} \vspace{1cm}\\
and \vspace{1cm}\\
A. Pashnev\footnote{BITNET: PASHNEV@THEOR.JINRC.DUBNA.SU}\vspace{1cm}\\
{\it JINR--Laboratory of Theoretical Physics} \\
{\it Dubna, Head Post Office, P.O.Box 79, 101 000 Moscow, Russia}
\vspace{1.5cm} \\
{\bf Abstract}
\end{center}
The superfield formulation of type II Green-Schwarz superstring
with $n= (1,0)$ worldsheet supersymmetry is constructed. It is shown
that the inclusion of the second spinor coordinate in the target
superspase leads to the possibility of the reparametrization invariant
description of the \sust in the absence of any field from the
two dimensional supergravity multiplet.
The twistor-like action of Chern-Simons type for bosonic string in
$D=3,4,6,10$ is deduced from the superstring action.

\begin{center}
{\it Submitted to Mod. Phys. Lett. A}
\end{center}
\vfill
\setcounter{page}0
\renewcommand{\thefootnote}{\arabic{footnote}}
\setcounter{footnote}0
\newpage
\section{Introduction}
In spite of the importance of the well-known $\kappa$ -symmetry of \supa
and \sust Lagrangians, its nature was not completely investigated up to now.
Therefore the idea of the reformulation of \supa and \sust Lagrangians
in which this symmetry has clear geometrical meaning is very attractive
and a great deal of attention has been devoted to this problem
\cite{STV}-\cite{DIS}.

For the first time this idea was realized by Sorokin, Tkach and Volkov
(STV) in the paper \cite{STV} in which the
twistor-like formulations for $D=3,4$ superparticle with local $n=1,2$
superconformal invariance were proposed. They further showed that the
fermionic $\kappa$-symmetry of the conventional Brink-Schwarz superparticle
is closely related to the local superconformal transformations in the
corresponding superworldline. The idea of double Grassmann analiticity
in proper and target superspaces was formulated
 and twistor-like actions for $D=3,4,6$ superparticle with
$n=1,2,4$ superconformal worldline symmetry was constructed in \cite{DS}
(including nontrivial background).

There has been previous work on the construction of the twistor-
like action for the $D=10,n=8$ superparticle \cite{{GHS},{DGS}} and, as it
was shown by Galperin and Sokatchev \cite{GS}, double
analiticity is not necessary and they constructed an action which has
similar form for all dimensions $D=3,4,6,10$ with $n=1,2,4,8$. This action
looks like a strightforward generalization of (STV) action for $D=3$ $n=1$
case \cite{STV}.

In all examples considered the $\kappa$-symmetry of the superparticle
was connected with the local superconformal transformations which form the
subgroup of the group of \sudi of corresponding \suli. Preliminary
investigations show that in the case of superstring the situation is
analogous. First steps in constructing the twistor-like form of
Green-Schwarz superstring \cite{GS} have been made in superconformal
gauge \cite{B}-\cite{VZ} and the problem of finding the reparametrization
invariant action was formulated. Partially this problem was solved for
heterotic \sust in different approaches\cite{T}-\cite{DIS}. However, all of
the actions in these papers contain some of the fields of two
dimensional supergravity multiplet to ensure the reparametrization
invariance of the action. If there are no additional fields, as in the paper
by Ivanov and Kapustnikov \cite{IK}, only one of two Virasoro conditions is
fulfilled and part of the reparametrization transformations is broken. In
\cite{DIS} it was shown how to include one additional field from
two dimensional supergravity multiplet to restore the reparametrization
invariance and twistor-like action with $n=(4,0)$ worldsheet \susy in $D=6$
was constructed.

In this paper we construct the twistor-like action for type II Green-Schwarz
superstring \cite{GS} with $n=(1,0)$ \sush, parametrized by two bosonic
(time-like and space-like) and one Grassmann coordinates. The inclusion
of second spinor Grassmann superfield, which is necessary to describe
type II superstring,
leads to rather nontrivial consequence: in spite of the absence of fields
from supergravity multiplet, the action is invariant under the general
reparametrizations of the \sush.

In the second part of the paper we consider simpler case of the
superparticle in the formulation of Sorokin, Tkach and Volkov\cite{STV}
and Galperin and Sokatchev\cite{GS}. We show in what sense this action is
invariant not only under the superconformal transformations, but under
the general reparametrizations of the $n$ -extended \suli as well.
The action for superstring and transformation laws for all superfields
under general reparametrizations of the \sush are constructed in the
third part of the work.

At the end of the paper we briefly discuss the twistor-like action for
bosonic string which is the part of the component action for type II
\sust. It has very simple form including only bilinear and trilinear
couplings and do not contain metric fields. Nevertheless,in $D=3,4,6,10$
this action is invariant under the transformations of two-dimensional
diffeomorphism group of the string worldsheet.

\section{The superparticle and its symmetry group}
In this section we briefly describe the Galperin-Sokatchev \cite{GS}
twistor-like formulation of $D=10$ superparticle action with manifest
$n=8$ world-line supersymmetry and show that the symmetry group of this
action actually includes the arbitrary reparametrizations of
the $(1,8)$ world-line superspace with 1 time-like even coordinate $\tau$
and $8$ odd coordinates $\eta^a$.
The Galperin-Sokachev action for $D=10$ superparticle
\be\label{2.1}
S=\int d\tau d^8\eta P_{a\mu}(D_aX^\mu -
iD_a\Theta\gamma^\mu\Theta) =\int d\tau d^8\eta P_{a\mu}\Omega_a^\mu.
\ee
is an integral over the $n=8$ world-line superspace and describes the
dynamics of $n=8$ superfields  $X(\tau,\eta)$ and $\Theta(\tau,\eta)$
which are correspondingly the commuting vector and anticommuting
Majorana-Weil spinor coordinates of $10$-dimensional target $n=1$
superspace. The Lagrange multiplier $P_{a\mu}$ is anticommuting
$10$-dimensional vector and carries also a world-line $O(8)$ index $a$.
The definition of $n=8$ supercovariant derivatives $D_a$ is as follows:
\be\label{2.2}
D_a=\frac{\partial}{\partial\eta_a} + i\eta_a\frac{\partial}{\partial\tau}
\;,\;\; \{D_a,\;D_b\}=2i\delta_{ab}\frac{\partial}{\partial\tau}.
\ee

As was shown in \cite{GS}, the action \p{2.1} is invariant under global
$N=1,\; D=10$ target-space supersymmetry and under the $n=8$ world-line
superconformal group which is the subgroup of the general
superdiffeomorphism group of the $(1,8)$ world-line superspace.

In spite of the fact that the action \p{2.1} is written in terms of the
flat supercovariant derivatives it is invariant under the arbitrary
reparametrization of the space $\tau,\eta_a$.
This miraculous fact for any $n$ is based on the following simple arguments.

The variation of the flat world-line supercovariant derivatives \p{2.1}
under the general infinitesimal transformation of world-line coordinates
\be\label{2.3}
\tau\rightarrow\tau + \alpha(\tau,\eta),\;\eta_a\rightarrow\eta_a +
\xi_a(\tau,\eta)
\ee
is nonhomogenious in the $D_a$ and contains the term with ordinary time
derivative
\be\label{2.4}
\delta D_a=-(D_a\xi_b)D_b + 2i\zeta_a\frac{\partial}{\partial\tau}
\ee
where
\be
\zeta_a=\xi_a +\frac{i}{2}D_a \Lambda, \;\;\; \Lambda\equiv \alpha +
i\eta_a\xi_a.
\ee
Only in the case of superconformal transformations for which $\zeta_a=0$
the last term in \p{2.4} disappears. However this term is not so dangerous.
To see this let us calculate the variation of the quantity $\Omega_a^\mu$
in \p{2.1}. The total variation of this quantity consists of two parts
$$
\delta\Omega_a^\mu=\delta^1\Omega_a^\mu+\delta^2\Omega_a^\mu
$$
first of which is due to variation \p{2.4} of flat supercovariant
derivatives \p{2.2} and second is due to variations of dynamical
superfields $\delta X^\mu \equiv {X'}^\mu(\tau',\eta')-X^\mu(\tau,\eta)$
etc. if we assume that these fields are not scalars under the general
coordinate transformations \p{2.3} . Simple calculations lead to
\be
\delta^1\Omega_a^\mu=-(D_a\xi_b)\Omega_a^\mu+2i\zeta_a(\dot{X}^\mu-
i\dot{\Theta}\gamma^\mu\Theta)
\ee
where $\dot{X}^\mu\equiv\frac{\partial}{\partial\tau}X^\mu$. Using the
identity
\be
D_b\Omega_a^\mu+D_a\Omega_b^\mu=-2iD_a\Theta\gamma^\mu D_b\Theta+2i\delta_
{ab}(\dot{X}^\mu-i\dot{\Theta}\gamma^\mu\Theta)
\ee
which is due to the commutation relations \p{2.2} and assuming the
following transformation laws for $X^\mu$ and $\Theta$
\bea\label{2.5}
\delta X^\mu &=& \zeta_aD_aX^\mu, \\
\delta\Theta &=& \zeta_aD_a\Theta,
\eea
we find the total variation of the $\Omega_a^\mu$
\be
\delta\Omega_a^\mu=\frac{i}{2}(D_aD_b\Lambda)\Omega_b^\mu+\zeta_bD_b
\Omega_a^\mu.
\ee
The linearity of this expression in $\Omega_a^\mu$ and $D_b\Omega_a^\mu$ is
crucial for invariance of the action \p{2.1}. Taking into account the
variation of the supervolume element
\be
\delta(d\tau d^n\eta)=(\frac{\partial}{\partial\tau}\alpha-\frac{\partial}
{\partial\eta_a}\xi_a)(d\tau d^n\eta)=(\dot{\Lambda}-D_a\xi_a)(d\tau d^n\eta)
\ee
and integrating by parts we find the following expression for $\delta S$:
\bea\label{2.6}
\delta S &=& \int d\tau d^n\eta\;\left\{ \;(\delta P_{a\mu}-\zeta_bD_bP_
{a\mu}-\frac{n-1}{2}\dot{\Lambda}P_{a\mu}-\right. \nn
 & & \\
 & & \left.
-\frac{i}{4}([D_a,D_b]\Lambda)P_{b\mu})\Omega_a^\mu-
 -D_b(\zeta_bP_{a\mu}\Omega_a^\mu)
\frac { } {}
\; \right\}\nonumber
\eea
Since the last term in the equation \p{2.6} is a total derivative, the
action is invariant if
\be\label{2.7}
\delta P_{a\mu}=\frac{n-1}{2}\dot{\Lambda}P_{a\mu}+\frac{i}{4}([D_a,D_b]
\Lambda)P_{b\mu}+\zeta_bD_bP_{a\mu}.
\ee
Thus the action \p{2.1} for any $n$ is also invariant under the arbitrary
reparametrizations of the $(1,n)$ superworldline . The transformation laws
\p{2.5} show that superfields $X^\mu(\tau,\eta)$ and $\Theta(\tau,\eta)$
which behaves as scalars under the transformations of superconformal group
lose this property for transformations \p{2.3} with $\zeta_a(\tau,\eta)\neq
0$. The transformation law for $P_{a\mu}$ (see \cite{GS}) is also modified by
the
last term in eq.\p{2.7} having the structure analogous to that in the
transformation laws for all other fields. The Chern-Simons nature of the
action \p{2.1} \cite {HT} , i.e. the invariance of the action under the
general coordinate transformations in the absence of one-dimensional
supergravity multiplet is confirmed.

The interesting feature of the transformation laws \p{2.5},\p{2.7} is that
in the active form (we consider for simplicity only $X^\mu$ )
\bea
\overline\delta X^\mu &\equiv& \delta X_\mu
-\xi_a \frac{\partial}{\partial\eta_a}X^\mu-
\alpha\frac{\partial}{\partial\tau}X^\mu  \nn
 &=& -(\Lambda-\frac{1}{2}\eta_aD_a\Lambda)\frac{\partial X^\mu}
{\partial\tau} +\frac{i}{2}D_a\Lambda \frac{\partial X^\mu}{\partial\eta_a}
\eea
they have the form of superconformal transformations of the scalar
super\-field \cite{GS}(appro\-pri\-ately modified for $P_{a\mu}$) with
parameter $\Lambda(\tau,\eta)$ constructed from arbitrary parameters
$\alpha(\tau,\eta)$ and $\xi_a(\tau,\eta)$ according to the eq. \p{2.4}.
Thus, all the superfields considered demonstrate some sort of mimicry,
i.e. their transformation laws \p{2.5},\p{2.7} for general
reparametrizations \p{2.3} of \sush are adapted in such a way, that only
particular combination $\Lambda$ of transformation parameters makes real
contribution to their active form of transformation laws. Analogous effect,
as we will see, takes place in the twistorlike formulation of \sust.

\setcounter{equation}0
\section{$n=(1,0)$ Superstring in $N=2$ extended target superspace}
The \sush of the \sust is parametrized by two bosonic $\tau^0,\tau^1$ and
one fermionic $\eta$ coordinates. We first introduce superfields
\bea
X^\mu(\tau^i,\eta) &=& x^\mu(\tau^i)+i\eta\chi^\mu(\tau^i),\\
P_\mu(\tau^i,\eta) &=& p_\mu(\tau^i)+i\eta\rho(\tau^i),\\
\Theta^A(\tau^i,\eta) &=&  \theta^A(\tau^i)+\eta\lambda^A(\tau^i),\;\;A=1,2,
\eea
and flat covariant derivative
\be
D=\frac{\partial}{\partial\eta}+i\eta\frac{\partial}{\partial\tau^0},\;\;
D^2=i\frac{\partial}{\partial\tau^0}.
\ee
in the superspace $(\tau_0,\eta)$.

We propose the following action for the \sust
\bea\label{Act}
S &=& \int d^2\tau d\eta \left\{ \left(-iP_\mu+\Theta^1\gamma_\mu
{\Theta^1}'-\Theta^2\gamma_\mu
{\Theta^2}'\right)\Omega^\mu_0
\right.\nn
 & & \left.+\left(D\Theta^1
\gamma_\mu\Theta^1-D\Theta^2\gamma_\mu\Theta^2\right)\Omega^\mu_1-\right.\\
 & & \;\;\;\;\;\;\left.-iD\Theta^1\gamma_\mu\Theta^1\Theta^2\gamma^\mu
{\Theta^2}'+i\Theta^1\gamma_\mu{\Theta^1}'
D\Theta^2\gamma^\mu\Theta^2\right\}\nonumber
\eea
where
\bea
\Omega^\mu_0&=&DX^\mu-iD\Theta^1\gamma^\mu\Theta^1-iD\Theta^2\gamma^\mu
\Theta^2\\
\Omega^\mu_1 &=& {X^\mu}'-i{\Theta^1}'\gamma^\mu\Theta^1-i{\Theta^2}'
\gamma^\mu\Theta^2,\\
 & & {\Theta^A}'\equiv \frac{\partial}{\partial\tau^1}\Theta^A.\nonumber
\eea

The action \p{Act} is invariant under the transformations of extended
$N=2$ \susy in target superspace $(X^\mu,\Theta^A)$
\be
\delta\Theta^A=\epsilon^A,\;\;\delta X^\mu=i\Theta^A\gamma^\mu\epsilon^A,
\;\;\delta P_\mu=0
\ee
in $D=3,4,6$ and $10$ where the relation
\be\label{G}
(\gamma^\mu)_{(\alpha\beta}(\gamma_\mu)_{\gamma)\delta}=0,
\ee
takes place. When the target superspace is not extended $(N=1)$, i.e.
$\Theta^2=0$, the quartic in $\Theta$'s terms disappear and to within the
redefinition of Lagrange multiplier $P_\mu$ our action
coinsides with the action for the $n=(1,0)$ twistorlike \sust proposed in
the paper \cite {DIS}(see also \cite{IK}). Such an extension of the
target superspace leads to highly nontrivial consequences. namely,
the action \p{Act} is invariant, as we will see later, under the general
reparametrizations of \sush $(\tau^i,\eta)$ without
introducing any additional fields such as supergravity multiplet. Due to
this invariance we can use usual (not lightlike) notations $\tau^0,\tau^1$
for bosonic coordinates of \sush .

The most general coordinate transformation of the \sush $(\tau^i,\eta)$
have the form:
\be\label{tr}
\delta\tau^i=\alpha^i(\tau,\eta), \;\; \delta\eta=\xi(\tau,\eta)
\ee
and the following combinations of these parameters
\be
\Lambda=\alpha^0+i\eta\xi, \;\; \zeta=\frac{i}{2}D\Lambda+\xi
\ee
will be useful as well. The transformation laws for derivatives and for the
integration measure are:
\bea\label{tr1}
\delta D &=& -(D\alpha^1)\frac{\partial}{\partial\tau^1}-(D\xi)D+2\zeta D^2\\
\delta\frac{\partial}{\partial\tau^1} &=& -{\alpha^1}'\frac{\partial}
{\partial\tau^1}-\xi' D+i\Lambda' D^2\\
\delta(d^2\tau d\eta) &=& ({\dot\alpha}^0+{\alpha^1}'-\frac{\partial\xi}
{\partial\eta})d^2\tau d\eta.
\eea

If the superfields $X^\mu,\Theta^A$ transform under the \p{tr} in the
following way
\bea\label{stl}
\delta X^\mu &=& \zeta DX^\mu+\frac{D\alpha^1}{2R^2}
(\Theta^1\gamma^\mu{\hat{P}\hat{R}}D\Theta^1+\Theta^2\gamma^\mu{\hat P}
{\hat R}D\Theta^2),\\
\delta\Theta^A &=& \zeta D\Theta^A-i\frac{D\alpha^1}{2R^2}
{\hat{P}\hat{R}}D\Theta^A\\
R_\mu &=& D \Theta^1\gamma_\mu D\Theta^1-D\Theta^2\gamma_\mu D\Theta^2\\
\delta P_\mu &=& \zeta DP_\mu-\frac{D\alpha^1}{R^2}
({\Theta^1}'\gamma_\mu\hat{P}\hat{R}D\Theta^1-{\Theta^2}'\gamma_\mu\hat{P}
\hat{R}D\Theta^2)-\nn
 & & \;\;\;-{\alpha^1}'P_\mu+\Lambda'R_\mu
\eea
where $\hat{P}\equiv P_\mu\gamma^\mu$, the integrand in \p{Act} is
invariant up to a total derivatives in critical dimensions $D=3,4,6,10$.
One can easily see that only two superfields $\Lambda$ and $\alpha^1$
contribute to the active form of these transformations in the whole
analogy with the case of superparticle.

To see that action \p{Act} indeed describes the superstring we
calculate the component form of the action. After performing the $\eta$-
integration we obtain
\bea\label{com}
S &=& \int d^2\tau \left\{ p_\mu(\omega_0^\mu-\lambda^1\gamma^\mu\lambda^1-
\lambda^2\gamma^\mu\lambda^2)+\omega_{1\mu}(\lambda^1\gamma^\mu\lambda^1-
\lambda^2\gamma^\mu\lambda^2)+\right. \nn
 & & \left.+i{\dot{x}}^\mu(\theta^1\gamma_\mu{\theta^1}'-\theta^2\gamma_\mu
{\theta^2}')
-i{x^\mu}'(\theta^1\gamma_\mu\dot\theta^1-\theta^2\gamma_\mu\dot\theta^2)+
\right. \nn
 & & \left.+\theta^1\gamma_\mu\dot\theta^1 \theta^2\gamma^\mu{\theta^2}'-
\theta^1\gamma_\mu{\theta^1}' \theta^2\gamma^\mu\dot\theta^2+\right.\\
 & & \left.+i(\rho_\mu+2\lambda^1\gamma_\mu{\theta^1}'-2\lambda^2\gamma_\mu
{\theta^2
}')(\chi^\mu-\lambda^1\gamma^\mu\theta^1-\lambda^2\gamma^\mu\theta^2)+
\right. \nn
  & & \left.+i(\chi^\mu(\lambda^1\gamma_\mu\theta^1-\lambda^2\gamma_\mu
\theta^2))'+
i(\lambda^1\gamma^\mu\theta^1 \lambda^2\gamma_\mu\theta^2)'\right\},\nonumber
\eea
where
\be
\omega_0^\mu=\dot{x}^\mu-i\dot\theta^1\gamma^\mu\theta^1-
i\dot\theta^2\gamma^\mu\theta^2,\;\;\omega_1^\mu={x^\mu}'
-i{\theta^1}'\gamma^\mu\theta^1-i{\theta^2}'\gamma^\mu\theta^2.
\ee

We first analize the structure of the action and investigate equations of
motion. The last two terms in the action \p{com} are total derivatives and
they give no contribution to the equations of motion. Only the third term
from the end of the action contains fields $\rho_\mu$ and $\chi^\mu$.
These fields are auxiliary because  their equations of motion simply express
them in terms of all other fields. The most fundamental role in the action
\p{com} is played by bosonic fields $\lambda^A$ which are spinors of the target
space. Their equations of motion
\bea
(\hat{p}-\hat{\omega}_1)\lambda^1 &=& 0\\
(\hat{p}+\hat{\omega}_1)\lambda^2 &=& 0
\eea
in dimensions $D=3,4,6,10$ lead to the following relations
\bea\label{l}
{p}_\mu-{\omega}_{1\mu} &=& e_1(\tau^i)( \lambda^1\gamma_\mu\lambda^1)\\
{p}_\mu+{\omega}_{1\mu} &=& e_2(\tau^i)(\lambda^2\gamma_\mu\lambda^2)
\eea
with two arbitrary functions $e_{1,2}(\tau^i)$. As a concequence, these
relations reproduce two Virasoro conditions
\be
(p_\mu\pm\omega_{1\mu})^2=0.
\ee
To construct the Lagrangian in more familiar form we first find expressions
$\lambda^A\gamma_\mu\lambda^A$ from \p{l}-(3.24) and substitute them in the
Lagrangian \p{com}. After elimination of $p_\mu$ due to it's equation
of motion, we will obtain the standard Lagrangian for superstring\cite{GS} in
which metric tensor $g^{ik}$ is expressed in terms of two independent
functions $e_{1,2}(\tau^i)$. Thus on mass shell the Lagrangian \p{com}
is equivalent to the Green-Shwarz Lagrangian\cite{GS}.
\setcounter{equation}0
\section{Discussion}
The twistor-like action \p{Act} is written in terms of $n=(1,0)$ worldsheet
superfields. In spite of the fact that only flat derivatives $D$ and
${\partial}/{\partial\tau^1}$ enter the formula \p{Act},
the action is
invariant under the general reparametrizations \p{tr} of the \sush
$(\tau^i,\eta)$. Unlike the previous investigations [6,15-17], we
consider the case of $N=2$ extended target superspace which correspond to the
type II Green-Schwarz superstring. In the limit when one of the Grassmann
superfields is zero, our action coincides with the action of the work
\cite{DIS} (see also \cite{IK} ). However the presence of second Grassmann
superfield makes it possible to wright down transformations for all
superfields under the arbitrary reparametrizations of superworldsheet
in absence of two dimensional supergravity multiplet.

To make things more simple, let us consider the twistor-like formulation
of bosonic string which is trivially obtained from the component action
\p{com} for superstring
\be\label{bAct}
S_B= \int d^2\tau \left\{ p_\mu(\dot{x}^\mu-\lambda^1\gamma^\mu\lambda^1-
\lambda^2\gamma^\mu\lambda^2)+{x^\mu}'(\lambda^1\gamma^\mu\lambda^1-
\lambda^2\gamma^\mu\lambda^2)\right\}.
\ee
The classical equivalence of the action \p{bAct} with the conventional action
for bosonic string can be proved along the same line, as it was proved for
the \sust in preceding section.
The action is written in terms of flat derivatives and contains only
bilinear and trilinear couplings.
In spite of the absence of any metric fields, this action is
reparametrization invariant with the following transformation laws for all
fields (in the active form):
\bea\label{tl}
{\overline\delta}\lambda^A &=& -\frac{1}{2}\dot\alpha^0\lambda^A+
\frac{\dot\alpha^1}{2r^2}\hat{p}\hat {r}\lambda^A\;
-\alpha^0\dot\lambda^A-\alpha^1{\lambda^A}',\\
r_\mu &=& \lambda^1\gamma_\mu \lambda^1-\lambda^2\gamma_\mu \lambda^2,\\
{\overline\delta}p_\mu &=& -{\alpha^1}'p_\mu+{\alpha^0}'r_\mu\;
-\alpha^0\dot{p}_\mu-{\alpha^1}{p_\mu}',\\
{\overline\delta}x^\mu &=& -\alpha^0{\dot x}^\mu-\alpha^1{x^\mu}'.
\eea
These transformation laws are simple consequences of superfield
transformation laws \p{stl} written in components.
The commutator  of two transformations \p{tl}-(4.5) has the same form
\be\label{comb}
[\overline\delta_1,\overline\delta_2]=\overline\delta_3
\ee
with parameters $\alpha^i_3$ expressed in terms of $\alpha^i_1$ and
$\alpha^i_2$ with the help of standard formulas of two dimensional
diffeomorphism group:
$$
\alpha^i_3=\alpha^k_1\partial_k\alpha^i_2-\alpha^k_2\partial_k\alpha^i_1.
$$
Note, that the closure property \p{comb} of transformations \p{tl}-(4.5)
takes place only in the target spaces with $D=3,4,6$ and $10$ in which
the relation  \p{G} is valid.

Only $x^\mu$ transform under the transformations of diffeomorphism
group as a scalar. The transformation laws for all other fields are more
complicated. In some sense these fields play the role of zweibein to
ensure the repapametrization invariance of the action.

In the case of \sust the situation is more complicated. Though the
transformations (3.15-3.18) compensate the arbitrary superworldsheet
reparametrization (3.10), their commutator in the active form contains
some additional transformations which also do not change the
superstring action \p{Act}. The description of the properties of
transformations (3.15)-(3.18) in more details will be given elsewhere.

\def\thesection { }
\section{Acknowledgements}

It is a pleasure for us to thanks participants of the seminar
``Supersymmetry-92" (1-3 of July 1992  in Dubna near the Moscow) for useful
discussions.

\end{document}